\documentclass[aps,preprint]{revtex4}

\usepackage{epsfig,amsmath}
\usepackage{subfigure}
\usepackage{graphicx}
\usepackage{dcolumn}
\usepackage{stmaryrd}
\usepackage{mathrsfs}
\usepackage{pifont}
\usepackage{amsthm}
\usepackage{amssymb}
\usepackage{bm}
\usepackage{latexsym}
\usepackage[colorlinks=true,linkcolor=blue,citecolor=blue]{hyperref}
\usepackage{color}

\newcommand{\la}{\langle}
\newcommand{\ra}{\rangle}

\newcommand{\Lam}{\Lambda}

\newcommand{\ga}{\gamma}
\newcommand{\Ga}{\Gamma}

\newcommand{\da}{\dagger}

\newcommand{\om}{\omega}

\newcommand{\non}{\nonumber}
\newcommand{\pa}{\partial}

\def\jpa#1{{ J.\ Phys.\ A} {\bf#1}}
\def\pra#1{{ Phys.\ Rev. A\/} {\bf#1}}
\def\prb#1{{ Phys.\ Rev. B\/} {\bf#1}}
\def\pre#1{{ Phys.\ Rev. E\/} {\bf#1}}
\def\prl#1{{ Phys.\ Rev.\ Lett.} {\bf#1}}
\def\sci#1{{ Science} {\bf#1}}
\def\annph#1{{ Ann.\ Phys.} {\bf #1}}
\def\pla#1{{ Phys.\ Lett. A\/} {\bf#1}}

\def\nat#1{{ Nature} {\bf#1}}

\def\jcp#1{{ J.\ Chem. \ Phys} {\bf#1}}
\def\pr#1{{ Phys.\ Rep.} {\bf#1}}
\def\njp#1{{ New\ J.\ Phys.} {\bf#1}}
\def\apl#1{{ Appl.\ Phys.\ Lett.} {\bf#1}}

\begin{document}

\title{Transient unidirectional energy flow and diode-like phenomenon induced by non-Markovian environments}

\author{Jun Jing$^{1,2}$, Dvira Segal$^{3}$, Baowen Li$^{4}$, and Lian-Ao Wu$^{2}$\footnote{Email address: lianao.wu@ehu.es}}

\affiliation{$^{1}$Institute of Atomic and Molecular Physics and Jilin Provincial Key Laboratory of Applied Atomic and Molecular Spectroscopy, Jilin University, Changchun 130012, Jilin, China \\ $^{2}$Department of Theoretical Physics and History of Science, The Basque Country University (EHU/UPV), PO Box 644, and Ikerbasque, Basque Foundation for Science, 48011 Bilbao, Spain \\ $^{3}$Chemical Physics Theory Group, Department of Chemistry, University of Toronto, 80 St. George St., Toronto, Ontario M5S 3H6, Canada \\ $^{4}$Department of Physics and Centre for Computational Science and Engineering, National University of Singapore, Singapore 117542, Republic of Singapore}

\date{\today}

\begin{abstract}
Relying on an exact time evolution scheme, we identify a novel transient energy transfer phenomenon in an exactly-solvable quantum microscopic model consisting of a three-level system coupled to two non-Markovian {\em zero-temperature} bosonic baths through two separable quantum channels. The dynamics of this model can be solved exactly using the quantum-state-diffusion equation formalism, demonstrating finite intervals of unidirectional energy flow across the system, typically, from the non-Markovian environment towards the more Markovian bath. Furthermore, when introducing a spatial asymmetry into the system, an analogue of the rectification effect is realized. In the long time limit, the dynamics arrives at a stationary state and the effects recede. Understanding temporal characteristics of directional energy flow will aid in designing microscopic energy transfer devices.
\end{abstract}


\maketitle

The analysis of simple-prototype quantum energy transfer problems~\cite{Qthermo,HRect,Rect} assists in elucidating fundamental thermodynamic concepts in open quantum systems~\cite{fund}. In the common construction, energy flow through a quantum system is generated by coupling it to two macroscopic objects, thermal reservoirs of different temperatures~\cite{bulk1,bulk2}. Alternatively, directional energy flow can be attained by supplying work into an asymmetric system~\cite{hanggi}. Energy transfer problems are interesting for exploring the foundation of classical statistical mechanics, quantum dynamics, and the crossover between the classical and quantum worlds~\cite{Zurek,Eisfeld}. For example, understanding the emergence of the Fourier's law of heat conduction from the principles of open quantum systems is a long standing problem~\cite{Fourier1,Fourier2,Fme,Fop,Fourier3}. Moreover, achieving control over energy flow is of an enormous importance in many areas of science and technology, including energy management in functional nanoscale devices~\cite{Hbump,Hflow,Geoflux,Pop}, realization of information processing and computation in open quantum systems~\cite{BWL}, control over molecular reactivity and dynamics~\cite{Miller}, and refrigeration in metal-superconductor junctions~\cite{Tannor}.

In microscopic devices, introducing a spatial asymmetry within an anharmonic structure can result in different magnitudes for the forward and backward currents, under the application of a reversed temperature bias~\cite{Casatirec}. This diode-like behavior had recently attracted considerable theoretical and experimental attention, including the demonstration of phononic~\cite{Zettl}, electronic~\cite{electronic1,electronic2,GiaE,GiaT} and photonic~\cite{photon,Casatirec,recMD4,recMD1,recMD2,recMD3,reviewrec} rectifications.

Given these developments, it is highly desirable to identify minimal conditions under which a diode-like behavior can be obtained, controlled, and enhanced~\cite{Rect,Miller}. Studies of rectification and unidirectional energy flow in quantum devices~\cite{HRect,Rect,Hflow} were typically performed under certain-standard approximations (semiclassical operation, neglecting coherences in the subsystem, assuming a unique steady state), adopting quantum master equation approaches. Few works had considered design principles on exactly-solvable quantum models \cite{Davies}. This problem is fundamentally important: {\em Can we derive, from microscopic quantum theories, sufficient or necessary boundary conditions for realizing a certain nonlinear energy transport in atomic, molecular or a nanoscale system?}

In this work, we aim at achieving unidirectional energy flow in an open quantum system. While our results correspond to a {\em temporal} behavior, they expose ingredients for asymmetric dissipation thus potentially, asymmetric nonlinear transport. We employ reservoirs with different (non-Markovian) spectral properties rather than with different temperatures. A structured-non-Markovian environment is characterized by the correlation timescale of its fluctuations~\cite{nonM}, while in a Markovian bath the memory time is shorter than any other characteristic timescale of the system of interest. The bath memory function dictates the manner in which information and energy flow from the system to the attached macroscopic bath, and the back-action of the bath on the system. A finite memory time is crucial for achieving control over the state of an open quantum system~\cite{RanCont}. Essentially, there is nearly no revival of the system's fidelity when it is attached to a memoryless Markovian bath~\cite{Minireview}. Therefore, if the system is coupled to two baths of different memory functions (correlation times, as defined below), a unidirectional flow can emerge: energy is fed back from the non-Markovian bath to the system, and simultaneously, the system is releasing its energy to the more Markovian bath.

We investigate the dynamics of our model by employing a nonperturbative master equation, derived from the quantum-state-diffusion (QSD) equation~\cite{QSD} (see \textbf{Method}). We show that due to the assignment of distinct memory properties to the baths, a transient unidirectional energy flow develops in a prototype model for energy flow across a quantum open system. The model consists a three-level system~\cite{Threelevel}, one of the simplest realizations of quantum engines~\cite{TransUT}, and two uncorrelated bosonic baths, bulk objects. Furthermore, we manifest that we can control the magnitude of the temporal energy flow by introducing a spatial asymmetry into the system, coupling it with different strengths to the contacts. This asymmetry grants an effect which can be categorized as a ``transient diode effect": the magnitude of the energy flow is different under forward and reversed operations, upon interchanging the channels connecting to the reservoirs.

\begin{figure}[htbp]
\centering
\includegraphics[width=3.2in]{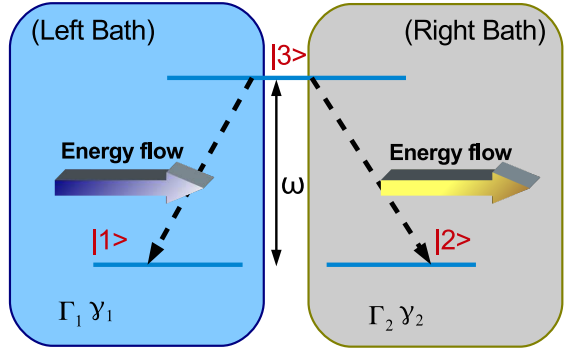}
\caption{Schematic representation of our model, a degenerate $\Lam$-type three-level system coupled to two zero-temperature uncorrelated baths characterized by different memory parameters $\ga$ and coupling constants $\Ga$.}\label{dia}
\end{figure}

\noindent \textbf{Results}\\
\textbf{Unidirectional energy flow.} To investigate the flow of energy in our system, purely induced by the distinct environmental memory functions, the two baths are assumed to be at zero-temperature and the three-level system is assumed to be of a degenerate $\Lambda$-type, where the energy splitting between the high level $|3\ra$ and the two lower levels, $|2\ra$ and $|1\ra$, is set as $\om$. The setup is shown in Fig.~\ref{dia}, where the energy current across the system is determined by the two energy flows between the system and the two baths. We use the following form for the correlation functions of the two baths, $\alpha_j(t,s)=\frac{\Ga_j\ga_j}{2}e^{-\ga_j|t-s|}$, with $j=1,2$, where $\Ga_j$ is the coupling strength of the system to the $j$th bath. This form corresponds to a Lorenz spectrum. When $\ga_j\rightarrow0$, the $j$th bath is eminently non-Markovian with long memory time as desired. On average, the dissipation rate of the system to the $j$th bath is greatly suppressed with decreasing $\ga_j$~\cite{QSD}. In contrast, when $\ga_j\rightarrow\infty$, the $j$th bath is memoryless. In this case, energy flow from the system to the bath is fully irreversible. Therefore, $1/\ga_j$ could be used to measure the environmental memory time.

\begin{figure}[htbp]
\centering
\subfigure{\label{EC1}\includegraphics[width=3in]{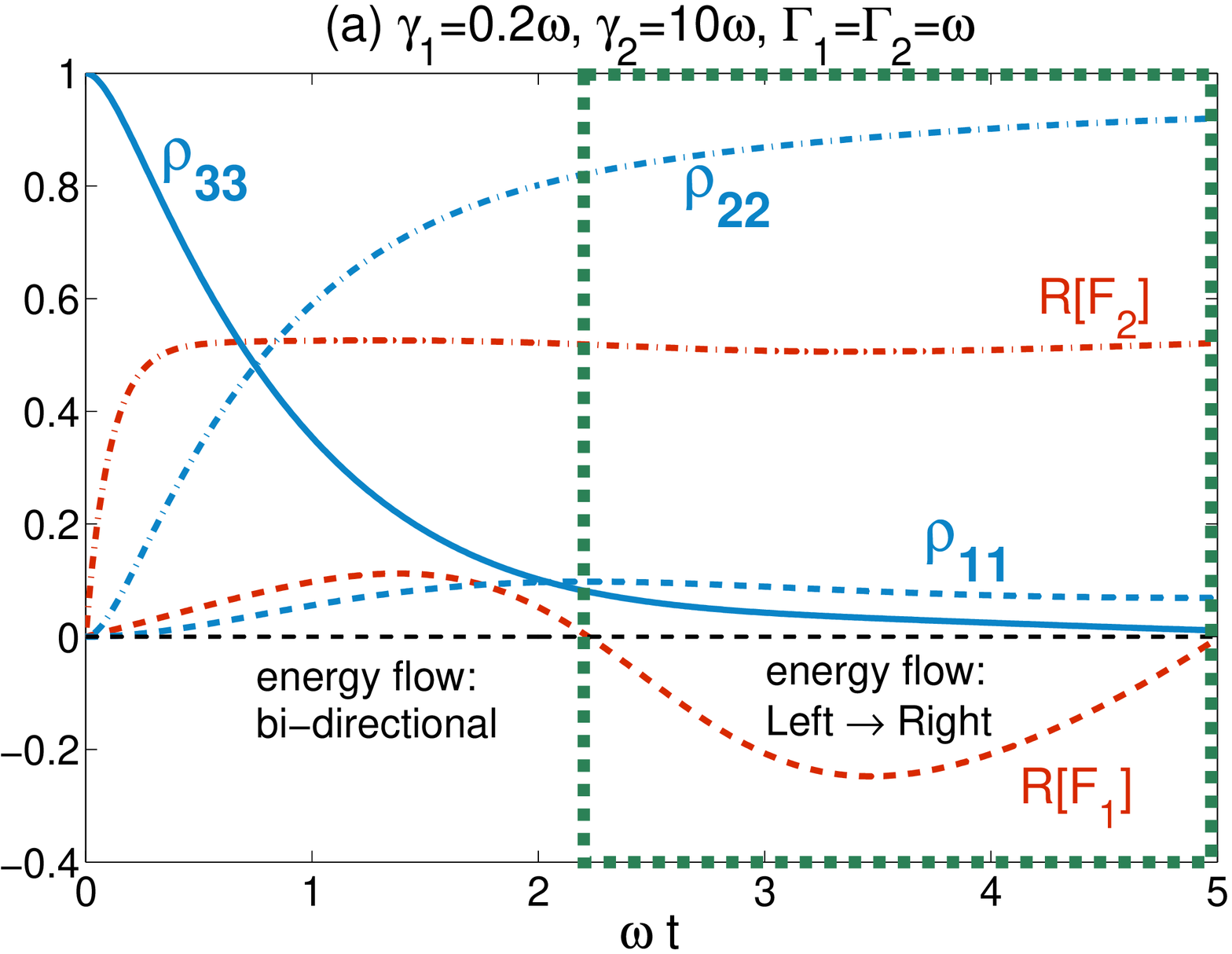}}
\subfigure{\label{EC2}\includegraphics[width=3in]{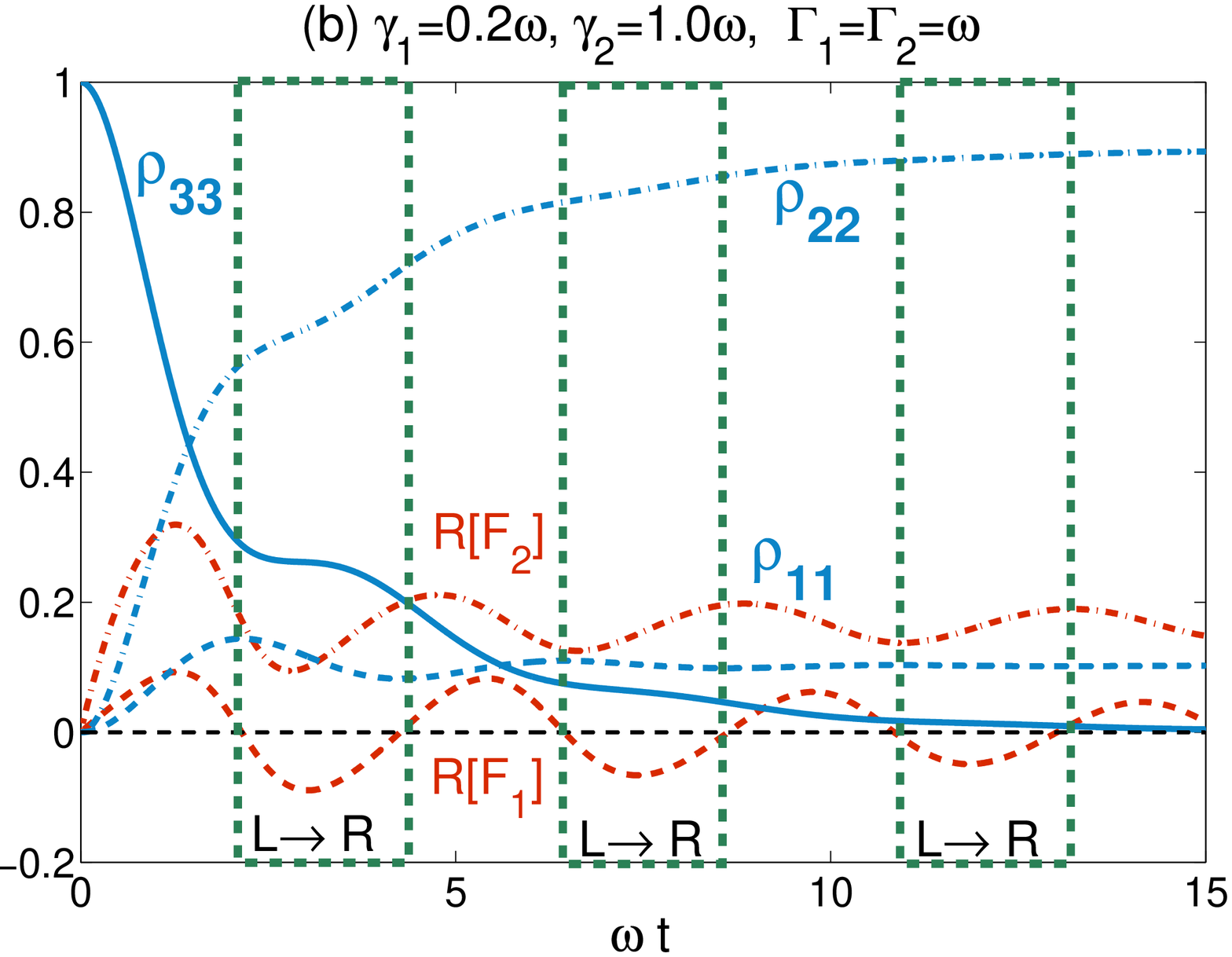}}
\subfigure{\label{Uni}\includegraphics[width=3in]{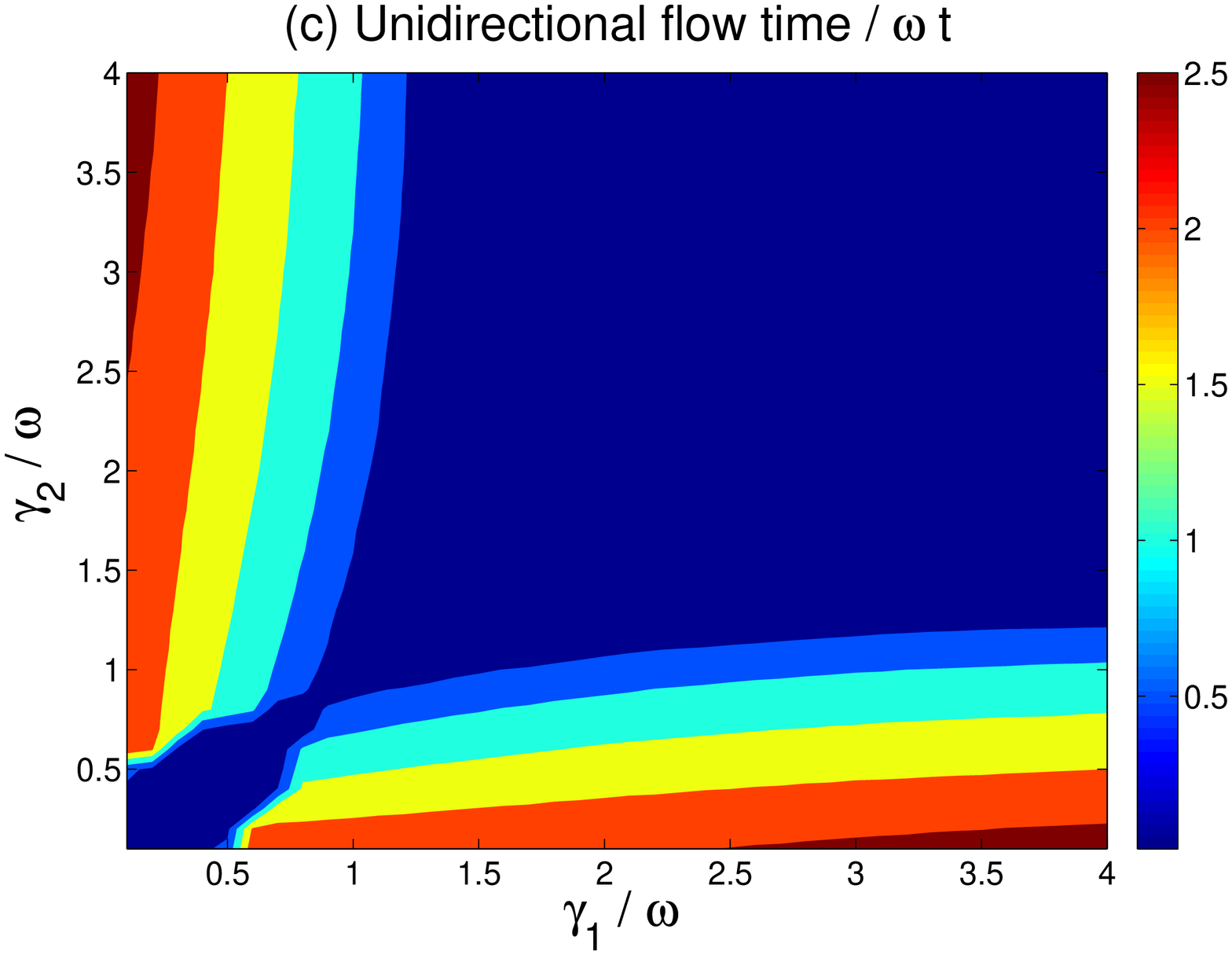}}
\caption{(Color online) (a) and (b) Population dynamics of the three states in the $\Lambda$ system, as well as our measure for energy flow, $\mathcal{R}[F_j(t)]$ with $\Gamma_1=\Gamma_2$, $\ga_1$= $0.2\om$ and different values for $\gamma_2$: (a) $\ga_2$=$10\om$, (b) $\ga_2$=$\om$. The $L\rightarrow R$ notation indicates that the unidirectional flow proceeds from the left bath to the right one within the marked frames}. (c) Duration time of unidirectional energy flow (in its first occurrence) as a function of the inverse memory time $\ga_1$ and $\ga_2$. The dark blue region corresponds to cases with vanishing unidirectional flow (zero duration).
\end{figure}

As our initial condition, we only excite state-$3$, $\rho_{33}(t=0)=1$. We demonstrate now the transient energy transfer behavior in the system when $\alpha_1(t,s)\neq\alpha_2(t,s)$, before reaching the stationary long time solution $\rho_{33}=0$. In Fig.~\ref{EC1} we use $\ga_1=0.2\om$ and $\ga_2=10\om$ and show that initially energy is released into both baths simultaneously. However, around $\om t=2.2$ the system begins to absorb energy from bath-$1$. As presented in \textbf{Method}, energy flow then becomes unidirectional, directed from the left reservoir (bath-$1$) towards the right side (bath-$2$), i.e., $\mathcal{R}[F_1]<0$ and $\mathcal{R}[F_2]>0$. We end our simulation when $\rho_{33}$ becomes extremely small and $\rho_{11}$ and $\rho_{22}$ reach the stationary states. The region of this unidirectional flow is embedded within a green-dashed frame.

The dynamics can be made more involved if both reservoirs are highly non-Markovian. In Fig.~\ref{EC2} we use $\ga_1=0.2\om$ and $\ga_2=1.0\om$, resulting in multiple-alternating regions of bi-directional and unidirectional energy flow. Particularly, we observe three intervals (distinguished by the green-dashed frames) of unidirectional flow of energy, of nearly the same duration yet shrinking amplitude, before full relaxation of the excited state is reached. As expected, the directional flow takes place from the reservoir with longer memory time, to the side with shorter memory time, since at zero-temperature a completely Markovian bath can only absorb energy. The comparison of Figs.~\ref{EC1} to \ref{EC2} also reveals that, as expected, in the latter case the total evolution time towards the stationary solution is longer than in the first case.

A finite difference between the memory parameters, $|\ga_1-\ga_2|\neq0$, is a necessary yet insufficient condition for the emergence of unidirectional flow of energy. This is shown in Fig.~\ref{Uni}, where we display the time duration of unidirectional flow in the first interval. Note that energy is flowing in opposite directions ($R\rightarrow L$ or $L\rightarrow R$) in the regions below and above the diagonal in Fig~\ref{Uni}. Recall that in Fig.~\ref{EC2} we show that there may be more than one occurrence of unidirectional transfer in the overall dynamics. We find that to observe the effect, it is necessary to employ a reservoir with a long memory time, for example, $\gamma_1/\omega<0.5$, and a second reservoir with a shorter memory time, $|\ga_1-\ga_2|/\omega\gtrsim0.5$.

\noindent \textbf{Diode-like phenomenon.}
So far, we have demonstrated that within a certain time interval energy may flow in a unidirectional manner only due to differences in the memory capabilities of the two reservoirs. In Figs.~\ref{EC1}, \ref{EC2} and \ref{Uni}, we used $\Ga_1=\Ga_2$, and it is obvious that the direction and magnitude of the flow will be fully reversed upon the interchange of the quantum channels indicated by $L_1$ and $L_2$, which is equivalent to an interchange in the values $\gamma_1$ and $\gamma_2$. However, distinct  memory times can {\em not} induce a diode-like phenomenon, an asymmetry in the magnitude of the unidirectional energy flow under opposite ``polarities", as we explain next.

\begin{figure}[htbp]
\centering
  \subfigure{\label{rec1}\includegraphics[width=3in]{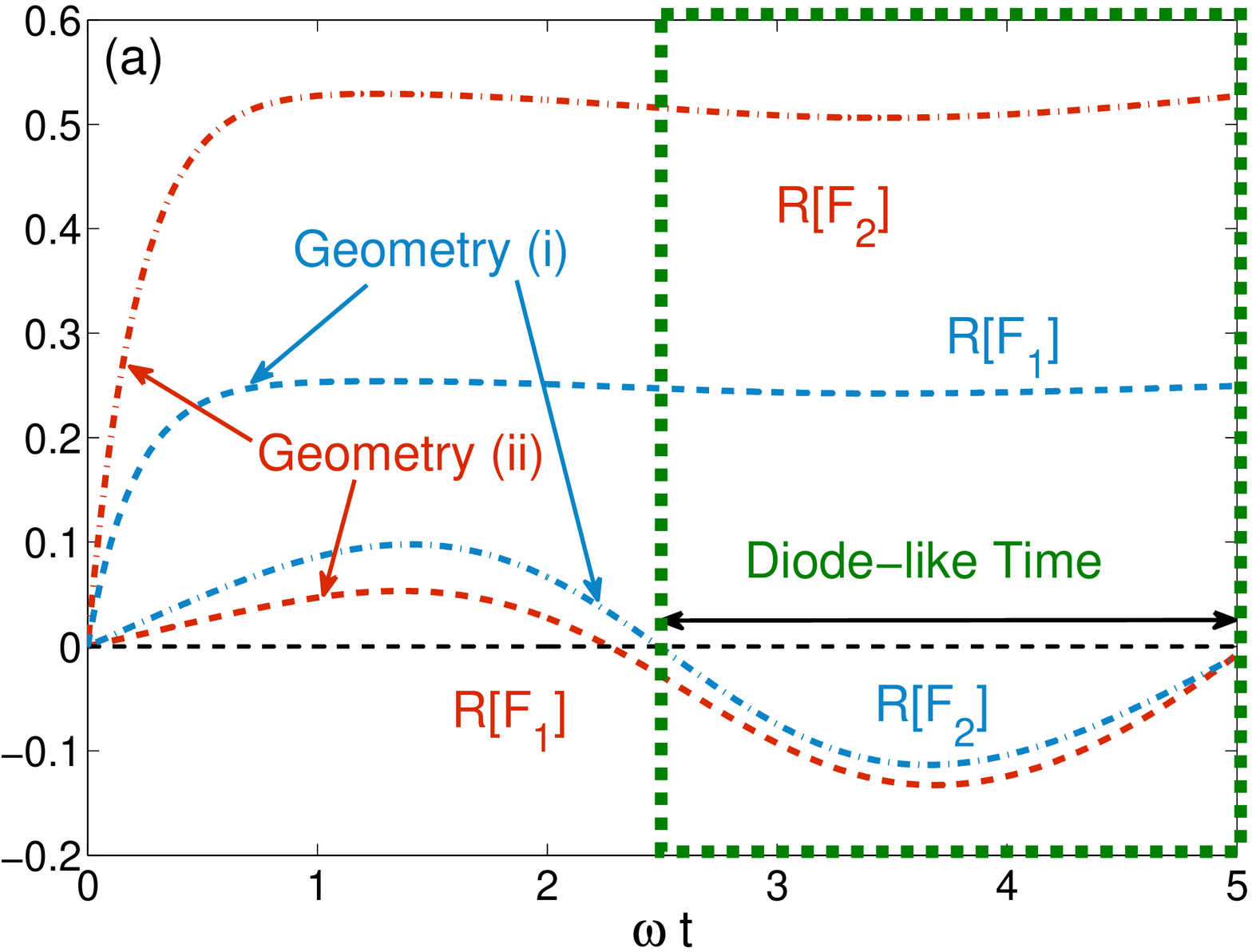}}
  \subfigure{\label{rec2}\includegraphics[width=3in]{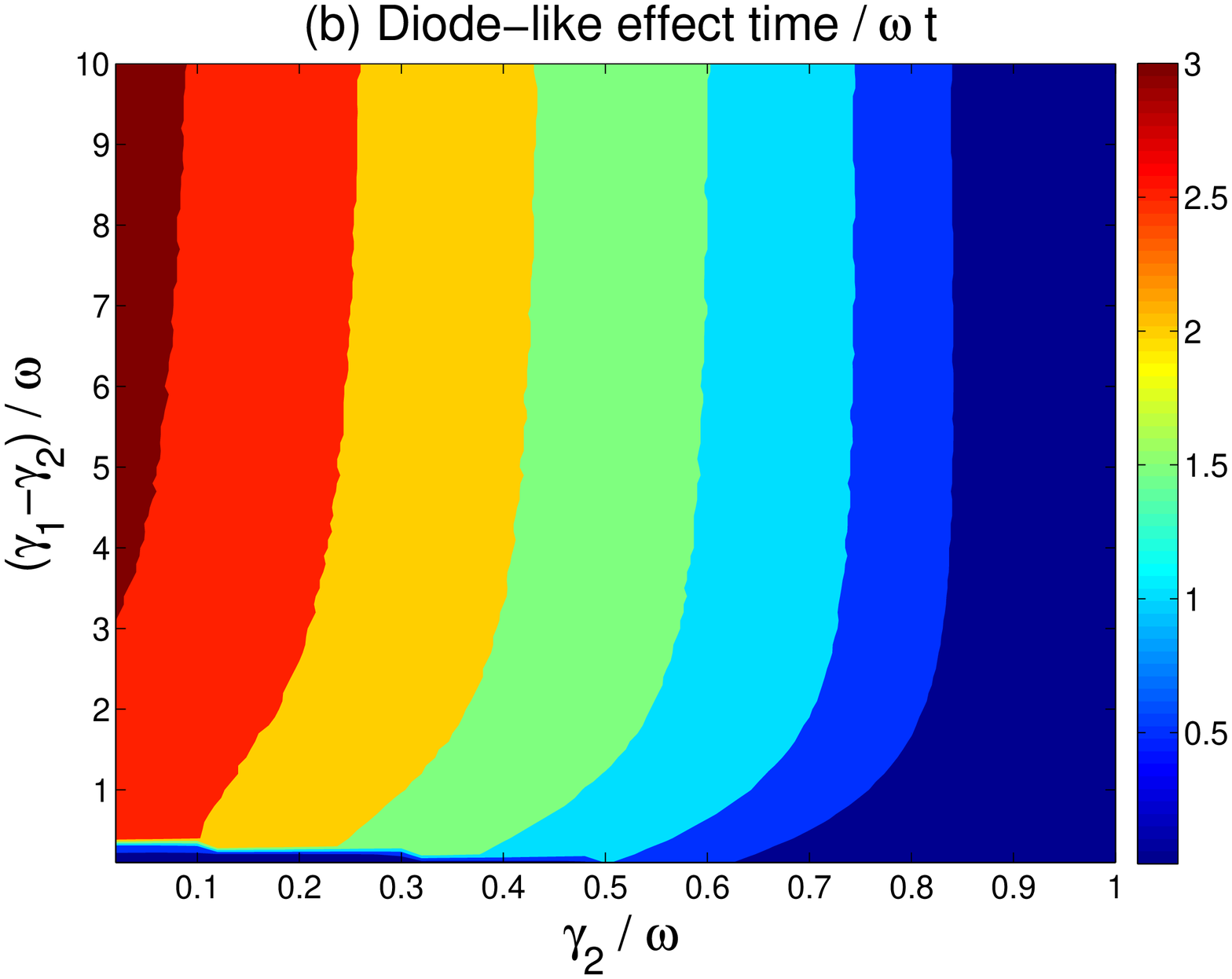}}
\caption{(Color online) (a) Transient unidirectional energy flow and a diode-like effect in two configurations: (i) (blue lines) $\gamma_1=5\omega>\gamma_2=0.2\omega$ and $\Ga_1=\Ga_2/2=\om/2$}, and (ii) (red lines) $\gamma_1=0.2\omega<\gamma_2=5\omega$ and $\Ga_1=\Ga_2/2=\om/2$. (b) Diagram of the diode-like time (the first period of unidirectional occurrence) in the parameter space of $\ga_1$ and $\ga_2$.
\end{figure}

We now show that the dynamics may be furthermore controlled by including an asymmetry in the coupling strengths of the three-level system to the baths, $\Gamma_1\neq\Gamma_2$. The resulting behavior corresponds to the thermal diode effect, see Figs.~\ref{rec1} and \ref{rec2}. We consider the following two setups: (i) a ``forward" configuration with $\gamma_1 > \gamma_2$ and $\Gamma_1<\Gamma_2$; (ii) a reversed geometry, in which we exchange the values of the memory times but keep the interaction energy as in (i), thus $\gamma_1< \gamma_2$ and $\Gamma_j$'s hold. We emphasize that the condition $\gamma_1 \neq \gamma_2$ allows a unidirectional flow of energy, and the spatial asymmetry $|\Ga_1-\Ga_2|\neq0$, provides the transient diode effect, yielding different magnitudes for energy flow in setups (i) and (ii).

We first examine geometry (i) in Fig.~\ref{rec1}. We use $\ga_1/\om=5>\ga_2/\om=0.2$ and $\Gamma_1=\Gamma_2/2$. Energy flows unidirectionally towards the more Markovian bath-$1$, as we found before, during the so called ``diode-like behavior interval" $2.5<\om t<5$. In geometry (ii) we employ $\ga_2/\om=5>\ga_1/\om=0.2$ while keeping $\Gamma_1=\Gamma_2/2$. Energy now flows (during almost the same interval) towards bath-$2$. In geometry (ii) the ratio of the energy flow towards the more Markovian bath of $\gamma=5\omega$ is larger than in case (i), given the stronger coupling to this bath. It is interesting to note that when we modify $\Gamma$ we largely affect the flow of energy into the more Markovian bath, as compared to changes in flow to the highly non-Markovian bath. We could explain this phenomenon by noting that $\Gamma$, the coupling strength of the system to the reservoirs, is the only parameter which determines the rate of energy flow to a Markovian bath. In contrast, the effectiveness of flow to a non-Markovian reservoir is predominated by the memory time of the bath, characterizing how effective it is in dissipating excess energy.

The duration of the diode-like behavior is plotted in Fig.~\ref{rec2} as a function of $\ga_1$ and $\ga_1-\ga_2$. The calculation is performed on a configuration similar to geometry (i) with $\ga_1$ taken always larger than $\ga_2$, providing a unidirectional flow from bath-$2$ to bath-$1$. We find that the interval of the diode effect (considering the first interval) is highly sensitive to the memory time of the non-Markovian bath, while the difference $\ga_2-\ga_1$ has a weaker overall effect. It is shown that a long diode-effect time is attainable when $\ga_2/\om\lesssim0.3$; the overall duration of the diode effect is fixed once $(\ga_1-\ga_2)/\om\gtrsim4$.

\noindent\textbf{Discussion} \\
We considered an exactly solvable model with two dissipation channels directed towards two reservoirs. We demonstrated that when adopting non-Markovian baths with different memory properties, the transient energy flow can become unidirectional, typically flowing from the highly non-Markovian to the more Markovian bath. This is the case as long as $|\ga_1-\ga_2|$ is larger than a certain threshold. Our analysis departs from the regular thermodynamic setup in which energy flow is driven by a temperature gradient across the system, to consider zero-temperature situations with non-Markovian baths. Moreover, we showed that the magnitude of the energy flow can be controlled to achieve an effect reminiscent of the diode phenomenon, by coupling the system to two contacts with different strengths. In conclusion, a sufficiently large difference in $|\ga_1-\ga_2|$ yields the effect of transient unidirectional energy flow; a sufficiently large difference in $|\ga_1-\ga_2|$, with the additional condition  of an asymmetrical coupling strengths between system and baths, results in a diode-like phenomenon.

Our $3$-level system could be realized in the triplet ground electronic state of a nitrogen vacancy (NV) center~{\cite{NV}}, when $\om$ is regarded as the zero field splitting. Transitions from the lower degenerate levels to the upper state can be selectively addressed via optical fields~\cite{Jelezko}. Besides optical relaxation, one of the two channels could be realized through the vibrations of the diamond lattice and the atoms comprising the NV center point defect. In this case, the photonic field would serve as the strong non-Markovian bath while the phononic environment would act as the more Markovian one.

Transient unidirectional flow of energy can be achieved in open quantum systems prepared in a non-stationary state, when coupled to two different structured environments. This behavior, obtained in our work by a microscopic quantum model without further assumptions and approximations, can be exploited for estimating the relative memory capabilities and non-Markovianity of competing baths, for constructing nonlinear quantum devices for the transport of energy, and for controlling unidirectional energy transfer, and potentially reactivity, in molecules~\cite{Dlott}. This is an exotic phenomenon induced by the boundary condition in quantum mechanics. Finally, the principles governing the dynamics of the present system at zero-temperature could be employed for exploring the dynamics of a finite-temperature, driven three-level system, to study the combined role of anharmonicity, non-Markovianity, driving, and asymmetry on energy transport phenomena~\cite{Goan,Roche}.

\noindent\textbf{Method}\\
The system takes a $\Lambda$-type configuration, with one excited state $|3\ra$ and two degenerate lower levels, $|1\ra$ and $|2\ra$. The excited state may decay to either of these lower levels, and these dissipation processes (referred as ``channels") are directed by two different baths~\cite{multichannel,comment1}. The baths are set at {\em zero-temperature}. The total Hamiltonian ($\hbar\equiv1$) is given by
\begin{equation}\label{eq:Ham}
H_{\rm tot}=\om|3\ra\la3|+\sum_{j=1,2;k}\om_{jk}a^\da_{jk}a_{jk}
+\sum_{j=1,2;k}\sqrt{\Ga_j}(g_ka_{jk}L_j^\da+g_k^*L_ja^\da_{jk}).
\end{equation}
Here, $\om$ is the energy splitting between the upper level and the lower two states. The system-environment coupling operators are represented by $L_j\equiv|j\ra\la3|$, where $j=1,2$. They open up energy transfer channels of the system into the $j$th bath. $a_{jk}^{\dagger}$ ($a_{jk}$) is the creation (annihilation) operator for independent mode-$k$ in the $j$th bath, $\sqrt{\Ga_j}g_k$ stands for the coupling constant between the system operator $L_j$ and the $k$th mode in the reservoirs. In our design unidirectional-transient flow is achieved by utilizing reservoirs with different two-time correlation functions $\alpha_j(t,s)$, defined below Eq.~(\ref{eq:psi}). Furthermore, when $\Gamma_1\neq\Gamma_2$, a process analogous to thermal rectification can be realized.

The wavefunction $|\Psi(t)\rangle$ denoted the solution to the Schr\"{o}dinger equation with the total Hamiltonian~(\ref{eq:Ham}) in the interaction picture with respect to $H_{\rm env}=\sum_{j=1,2;k}\om_{jk}a^\da_{jk}a_{jk}$. We define $|\psi_t(z^*)\rangle=\langle z_1|\langle z_2|\Psi(t)\rangle$ where $|z_j\rangle=\Pi_k^\otimes|z_{jk}\rangle$ stands for the tensor product of the Bargmann coherent states for the environment modes. The exact QSD equation~\cite{QSD}, for the stochastic wave-function $|\psi_t(z^*)\rangle$, is given as
\begin{equation}\label{eq:psi}
\pa_t|\psi_t(z^*)\ra=\left\{-i\om|3\ra\la3|+ \sum_{j=1,2}\left[L_jz_{jt}^*-F_j(t)|3\ra\la3|\right]\right\}|\psi_t(z^*)\ra.
\end{equation}
Here $z_{jt}^*\equiv-i\sqrt{\Ga_j}\sum_{k}g_k^*z_{jk}^*e^{i\om_{jk}t}$ are correlated processes, describing the stochastic influence of the $j$th bath. $z_{jk}^*$ are individual Gaussian-distributed complex random variables, whose ensemble average is defined as $M[z_{jt}z_{js}^*]=\alpha_j(t,s)=\Ga_j\sum_k|g_k|^2e^{-i\om_{jk}(t-s)}$. The operator $F_j(t)|3\ra\la3|$ includes the effect of the $j$th bath on the system dynamics. The function $F_j(t)\equiv\int_0^tds\alpha_j(t,s)f_j(t,s)$ satisfies
\begin{equation}\label{eq:fts}
\pa_tf_j(t,s)=[i\om+F_1(t)+F_2(t)]f_j(t,s),
\end{equation}
with $f_j(t,t)=1$ and $j=1,2$. The corresponding exact master equation for the reduced density matrix $\rho(t)=M[|\psi_t(z^*)\ra\la|\psi_t(z^*)|]$ can be constructed via the Novikov theorem~\cite{YUPRA99},
\begin{equation}
\pa_t\rho=[-i\om|3\ra\la3|, \rho]+\sum_{j=1,2}\left\{F_j(t)[L_j\rho, L_j^\da]+h.c.\right\}. \label{eq:DM}
\end{equation}
Equation~(\ref{eq:DM}) immediately yields the solution of time-dependent populations of the three levels,
\begin{eqnarray}\non
\rho_{33}(t)&=&e^{-\bar{F}_1(t)-\bar{F}^*_1(t)-\bar{F}_2(t)
-\bar{F}^*_2(t)}\rho_{33}(0),\\ \non
\rho_{22}(t)&=&\int_0^t[F_2(s)+F_2^*(s)]\rho_{33}(s)ds+\rho_{22}(0), \\
\rho_{11}(t)&=&\int_0^t[F_1(s)+F_1^*(s)]\rho_{33}(s)ds+\rho_{11}(0),
\end{eqnarray}
where $\bar{F}_j(t)\equiv\int_0^tdsF_j(s)$. Since the system is prepared in its excited state, $\rho_{33}(0)=1$ and $\rho_{22}(0)=\rho_{11}(0)=0$, the time-dependent energy current, defined positive when flowing from the system towards bath-$1$ and bath-$2$, is given by $2\om\mathcal{R}[F_1(t)]\rho_{33}(t)$ and $2\om\mathcal{R}[F_2(t)]\rho_{33}(t)$, respectively, where $\mathcal{R}[\cdot]$ stands for real part. These expressions identify the current as the population relaxation rate times the energy difference for the transition. For example, state-$3$ decays to state-$1$ by giving up energy through channel-$1$ to bath-$1$. Population decays from level-$3$ to level-$1$ thus directly relates to the amount of energy flowing from the system to the attached bath. The transitory energy transfer can therefore be measured by the real part of dimensionless coefficients $F_j(t)$, $j=1,2$, without invoking confusion, before $\rho_{33}$ vanishes.

Placing bath-$1$ ($2$) at the left (right) side of the system, we now identify different transport situations: (A) when $\mathcal{R}[F_1(t)]>0$ and $\mathcal{R}[F_2(t)]>0$, the three-level system is releasing energy to both sides; (B) the system is releasing energy to the left while absorbing energy from the right side when $\mathcal{R}[F_1(t)]>0$ and $\mathcal{R}[F_2(t)]<0$; (C) energy flows towards the right bath in the opposite scenario, $\mathcal{R}[F_1(t)]<0$ and $\mathcal{R}[F_2(t)]>0$; (D) the system is absorbing energy from both reservoirs at the same time if $\mathcal{R}[F_1(t)]<0$ and $\mathcal{R}[F_2(t)]<0$. Manifesting the development of scenarios (B) and (C), with a (transient) unidirectional energy flow as indicated by $\mathcal{R}[F_1(t)]\mathcal{R}[F_2(t)]<0$, is the objective of our work. If the memory functions have the same spectral form, $\alpha_1(t,s)/\Ga_1=\alpha_2(t,s)/\Ga_2$, then $F_1(t)/\Ga_1=F_2(t)/\Ga_2$ from Eq.~(\ref{eq:fts}). As a result, $\mathcal{R}[F_1(t)]$ and $\mathcal{R}[F_2(t)]$ acquire the same sign at all times, thus a unidirectional energy flow across the system cannot be realized, even when $\Ga_1\neq \Ga_2$. Evidently, to achieve our goal we should employ reservoirs with distinct memory properties.

\textbf{Acknowledgements}\\
We acknowledge grant support from the Basque Government (grant IT472-10), the Spanish MICINN (No. FIS2012-36673-C03-03), the NSFC No. 11175110, and Science and Technology Development Program of Jilin Province of China (20150519021JH). DS acknowledges support from an NSERC discovery grant and the Canada Research Chair Program. BWL acknowledges support from the Ministry of Education, Singapore, by Grant No. MOE2012-T2-1-114.

\textbf{Author contributions}\\
J.J. performed numerical simulations, analyzed results, and prepared figures. L.-A.W. contributed to the conception and development of the research problem. All authors (J.J., D.S., B.L. and L.A.W.) discussed the results and physical implications, and wrote the manuscript.

\textbf{Additional Information} \\
Competing financial interests: The authors declare no competing financial interests.


\begin{thebibliography}{99}
\bibitem{comment1}
In our model, the two dissipation channels correspond to different bath operators thus cross correlations of the form $M[z_{jt}z_{j's}^*]=\alpha_{j,j'}(t,s)$ are missing. The case in which all dissipation channels couple to the same bath is considered in Ref.~\cite{multichannel}.
\end{thebibliography}
\end{document}